\documentclass{article}

\usepackage{arxiv}

\usepackage{amsmath}
\usepackage[utf8]{inputenc} 
\usepackage[T1]{fontenc}    
\usepackage{hyperref}       
\usepackage{url}            
\usepackage{booktabs}       
\usepackage{amsfonts}       
\usepackage{nicefrac}       
\usepackage{microtype}      
\usepackage{cleveref}       
\usepackage{lipsum}         
\usepackage{graphicx}
\usepackage{natbib}
\usepackage{doi}
\usepackage{multirow}

\title{SNR-based beaconless multi-scan link acquisition model with vibration for LEO-to-ground laser communication}

\newif\ifuniqueAffiliation
\uniqueAffiliationtrue

\ifuniqueAffiliation 
\author{ \href{https://orcid.org/0000-0002-2471-1574}{\includegraphics[scale=0.06]{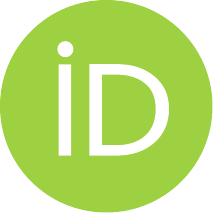}\hspace{1mm}Sen Yang} \\
	School of Astronautics and Aeronautics\\
	University of Electronic Science and Technology of China\\
	Chengdu, China 611731 \\
	\texttt{yang\_jansen@163.com} \\
	\And
	\href{https://orcid.org/0000-0000-0000-0000}{\includegraphics[scale=0.06]{orcid.pdf}\hspace{1mm}Xiaofeng Li} \\
	School of Astronautics and Aeronautics\\
	University of Electronic Science and Technology of China\\
	Chengdu, China 611731 \\
	\texttt{lxf3203433@uestc.edu.cn} \\
}
\else
\usepackage{authblk}

\newbox{\orcid}\sbox{\orcid}{\includegraphics[scale=0.06]{orcid.pdf}} 
\author[1]{%
	\href{https://orcid.org/0000-0002-2471-1574}{\usebox{\orcid}\hspace{1mm}Sen Yang\thanks{\texttt{yang\_jansen@163.com}}}%
}
\author[1]{%
	\href{https://orcid.org/0000-0000-0000-0000}{\usebox{\orcid}\hspace{1mm}Xiaofeng Li\thanks{\texttt{lxf3203433@uestc.edu.cn}}}%
}
\affil[1]{School of Astronautics and Aeronautics, University of Electronic Science and Technology of China, Chengdu, China 611731}
\fi

\renewcommand{\shorttitle}{\textit{arXiv} Template}

\hypersetup{
pdftitle={School of Astronautics and Aeronautics\\
	University of Electronic Science and Technology of China\\
	Chengdu, China 611731 \\},
pdfsubject={JASR},
pdfauthor={Sen Yang, Xiaofeng Li},
pdfkeywords={Transmitted power, Signal-to-noise ratio (SNR), Platform vibration, Multi-scan acquisition model, Optimizations},
}

\begin{document}
\maketitle

\begin{abstract}
We propose a link acquisition time model deeply involving the process from the transmitted power to received signal-to-noise ratio (SNR) for LEO-to-ground laser communication for the first time. Compared with the conventional acquisition models founded on geometry analysis with divergence angle threshold, utilizing SNR as the decision criterion is more appropriate for practical engineering requirements. Specially, under the combined effects of platform vibration and turbulence, we decouple the parameters of beam divergence angle, spiral pitch, and coverage factor at a fixed transmitted power for a given average received SNR threshold. Then the single-scan acquisition probability is obtained by integrating the field of uncertainty (FOU), probability distribution of coverage factor, and receiver field angle. Consequently, the closed-form analytical expression of acquisition time expectation adopting multi-scan, which ensures acquisition success, with essential reset time between single-scan is derived. The optimizations concerning the beam divergence angle, spiral pitch, and FOU are presented. Moreover, the influence of platform vibration is investigated. All the analytical derivations are confirmed by Monte Carlo simulations. Notably, we provide a theoretical method for designing the minimum divergence angle modulated by the laser, which not only improves the acquisition performance within a certain vibration range, but also achieves a good trade-off with the system complexity.
\end{abstract}

\keywords{Transmitted power \and Signal-to-noise ratio (SNR) \and Platform vibration \and Multi-scan acquisition model \and Optimizations}

The demand for larger bandwidth in modern satellite communication to handle vast amounts of data has rendered traditional radio frequency link with its low bandwidth and slow modulation rate impractical \citep{toyoshima2007comparison}. Free-space optics communication (FSOC), which offers several benefits such as a high bandwidth, use of a license-free spectrum, low power, and small form factor requirements, proves to be an excellent solution \citep{toyoshima2005trends}. Over the past few decades, numerous missions have yielded valuable achievements and catalyzed technological advancements. Such as low-Earth orbit (LEO) to ground laser communication experiment of the STRV-2 module in 2000 \citep{kim2001lessons}, a repeatable 5.625 Gbps bidirectional laser communication at 1064 nm between the NFIRE satellite and an optical ground station \citep{fields20115}, and a high-performance laser communication terminal developed by TESAT that fulfills the need of a power efficient system with a homodyne detection scheme and a BPSK modulation format \citep{gregory2017tesat}.

The acquisition, pointing, and tracking (APT) system plays a crucial role in establishing a stable FSOC link between two terminals \citep{young1986pointing}. Some FSOC systems \citep{picchi1986algorithms,yu2017theoretical,hu2022multi} employ the beacon strategy, which is commonly composed of two steps. First, the coarse APT emitting a beacon light with a large divergence angle and sufficient peak power is carried out to achieve a rough line-of-sight (LoS) alignment between the transmitter and receiver. Next, the transmitter employs a separate beam with a narrow divergence angle to enhance the alignment for fine APT \citep{ho2007pointing}. The disadvantage of this beacon-based strategy is that it requires additional beacon laser, resulting in the large scale of FSOC system. Beaconless FSOC system has been proposed to simplify terminal structure and reduce power while maintaining performance compared to classical beacon-based strategy, where a narrow single beam is adopted both for APT and data transmission \citep{hindman2004beaconless}. Subsequently, successful beaconless satellite-to-ground communication links were established using a compatible ground terminal \citep{gregory2017tesat}, and operational considerations for the beaconless spatial acquisition were presented in \citet{sterr2011beaconless}. However, the acquisition process poses significant challenges due to the narrow beam divergence angles and random vibration disturbances \citep{ho2007pointing}. To address these challenges, analytic expressions and optimizations for multi-scan average acquisition time were presented in \citet{li2011analytical}, taking into account factors such as the initial pointing error, beam divergence angle, and field of uncertainty (FOU). In addition, an approximate mathematical model was established in \citet{friederichs2016vibration} to describe the influence of Gaussian random vibration on the acquisition probability. Furthermore, \citet{ma2021satellite} derived an approximate analytical expression for the scan loss probability, considering the scanning parameters and platform vibrations, whose influence on acquisition time was analyzed under both single-scan and multi-scan patterns.

The mathematical models mentioned above assume that a successful acquisition requires the receiver to be within the beam divergence angle. They mainly focused on the first phase of the acquisition process, specifically on the scanning with which the beam enters the receiver antenna, lacking in-depth study on whether the optical signal incident on the photodetector can be effectively responded. In the APT system, photodetectors, such as four-quadrant detector of position sensors and charge-coupled device of image sensors, are utilized to correct the deviation of output spot \citep{qiu2021active,yang2022iterative}. However, the sensor still has output even without beam incidence, which is caused by noise such as dark current \citep{li2015limited}. If the response threshold is set too low, the noise will be misjudged as an optical signal, resulting in a false alarm. Conversely, If the threshold is set too high, the signal will be misjudged as noise, resulting in a missed detection. However, the absolute thresholds are different for types of sensors involving materials and other factors. Hence, signal-to-noise ratio (SNR) as a relative value becomes an appropriate indicator for criterion. Moreover, the optical intensity is also affected by the atmospheric turbulence in the satellite-to-ground laser communication \citep{kaushal2016optical}, where SNR is further reduced. Therefore, the derivations only based on geometric analysis are not rigorous.

\citet{hechenblaikner2023optical} investigated the acquisition performance with the determined power incident on the photodetector. It is a significant exploration, but the key parameter of transmitted power was not studied in depth in the model, which has the most direct and complete end-to-end relationship with the received SNR, as well as directly affects the complexity of the terminal structure. For LEO-to-ground \citep{hemmati2020near} FSOC with a narrow acquisition window, the scanning needs to maintain the maximum transmitted power to achieve fast acquisition. However, there is no theoretical derivation and optimization of acquisition time model involving the entire process from the transmitted power to received SNR to the best of our knowledge. Spurred by the gap, the coverage factor, representing the ratio of the range of a beam to the spiral pitch wherein the received SNR over the threshold at a certain transmitted power, is first defined in Section \ref{sec:2}. Then the power model is developed as a function of the beam divergence angle, the spiral pitch, and the coverage factor. Subsequently, we derive the probability distribution of the coverage factor, further combined with FOU and receiver field angle, we calculate the single-scan acquisition probability. In Section \ref{sec:3}, given the essential reset time between single-scan, we establish a novel multi-scan acquisition time model. By utilizing the decoupled relationship between the beam divergence angle and the spiral pitch, the optimizations concerning the acquisition parameters, also including FOU, are presented. Moreover, the influence of platform vibration is investigated. Finally, Monte Carlo (MC) simulations are performed in Section \ref{sec:4} to verify the above theoretical derivations and optimization conclusions.

\section{Single-scan mathematical model}\label{sec:2}
\begin{figure}[htpb]
	\centering\includegraphics[width=1.0\columnwidth]{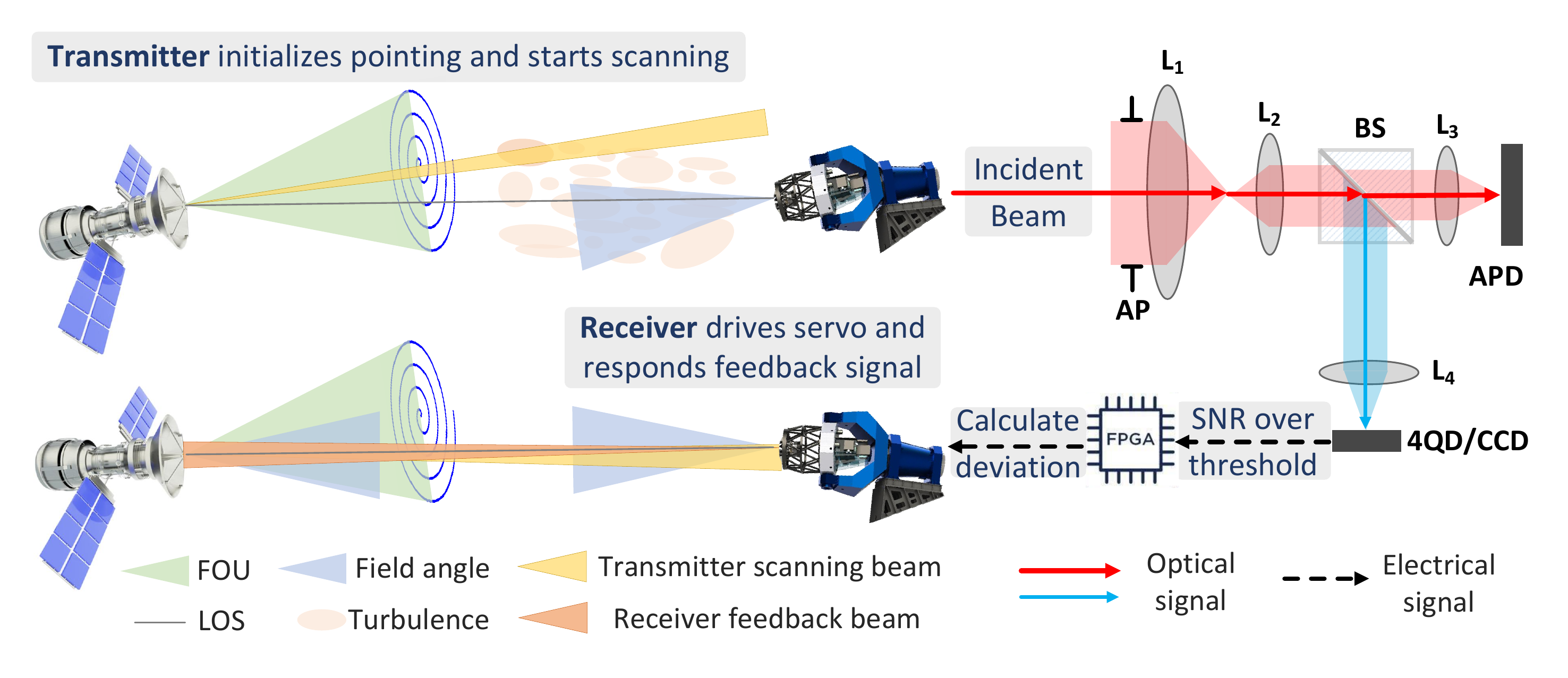}
	\caption{The diagram of beaconless acquisition. The error arises between the satellite initial pointing (the center of the spiral) and the LoS. The LoS is covered by a green conical shape, which allows the beam propagation towards the receiver antenna through scanning. Similarly, the blue conical shape covers the LoS, enabling the received photons of the signal beam to fall on the photodetector. The incident beam passes through an AP and is then expanded by the telescope system $L_1-L_2$. It subsequently passes through a BS and most of it is focused on an APD through $L_3$ for signal processing. The remaining small portion of the beam is detected by a 4QD or CCD through $L_4$ for spot deviation measurement. $L_1,L_2,L_3,L_4$, lens; AP, aperture; BS, beam splitter; APD, avalanche photo diode; 4QD, four-quadrant detector; CCD, charge-coupled device photodetector.}
	\label{fig:1}
\end{figure}

During the establishment of a space laser link with beaconless APT, the signal beam divergence angle is markedly small, usually measured in microradians. While the discrepancy arising between the transmitter initial pointing and the LoS typically ranges in the magnitude of milliradians and exhibits a random distribution due to the accuracy errors from the satellite attitude, orbit prediction, and terminal control \citep{gao2023improved}, resulting in the uncertainty of beam pointing. Consequently, the Archimedes spiral technique \citep{steinhaus1999mathematical} is commonly adopted for scanning the FOU. While the receiver keeps staring at the transmitter. Once the receiver detects a laser signal that satisfies the specified SNR requirements, it calculates the spot deviation using the quantized response from the photodetector. The receiver then drives the high-precision servo mechanism to achieve a fine tuning on the pointing \citep{riel2020high}, and responds with an optical signal to stimulate the response of the photodetector at the transmitter, thereby completing the acquisition process. The acquisition diagram is depicted in Fig. \ref{fig:1}.

\subsection{Power model}\label{sec:2.1}
The average SNR of FSOC system with intensity modulation / direct detection scheme is \citep{andrews2005laser}:
\begin{eqnarray}
	\bar{Q}={\left\langle i_{s}^{2} \right\rangle }/{\left\langle i_{n}^{2} \right\rangle }
	\label{eq:1}
\end{eqnarray}
where $i_{s}$ is signal current, ${i}_{s}={P}_{r}{R}_{r}={P}_{t}{h}_{t}{h}_{c}{R}_{r}$, $P_{t}$ and $P_{r}$ are the transmitter and receiver powers, respectively, $h_t$ represents the transmission gain with vibration, $h_c$ represents the turbulence attenuation, the two are independent \citep{jurado2012impact}, and ${R}_{r}$ is photoelectric response efficiency. $i_{n}$ is noise current, which is additive Gaussian white noise with zero mean and $\sigma _{n}^{2}={{N}_{0}}$ variance. Therefore, Eq. (\ref{eq:1}) is specifically expressed as:
\begin{eqnarray}
	\bar{Q}={E\left[ h_{t}^{2} \right]E\left[ h_{c}^{2} \right]P_{t}^{2}R_{r}^{2}}/{{{N}_{0}}}
	\label{eq:3}
\end{eqnarray}

\begin{figure}[htbp!]
	\centering\includegraphics[width=0.8\columnwidth]{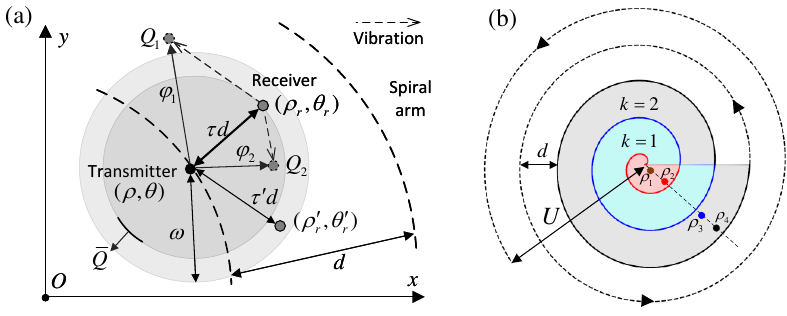}
	\caption{Scanning diagram. (a) Details for scanning. $d$ is spiral pitch, and $\omega $ is beam divergence angle. $\tau d$ is the minimum distance between the receiver $({\rho }_{r},{\theta }_{r})$ and the spirals. The platform vibration from transmitter is equivalent to the deviation on the receiver, so that ${{\varphi }_{n}}(n=1,2,\ldots )$ are random variables representing the angle deviation between the transmitter pointing and LoS. At a certain transmitted power, the corresponding instantaneous SNR are ${{Q}_{n}}$, and the received average SNR is equal to the threshold $\bar{Q}$, which is the lowest value in a circular region with transmitter $(\rho, \theta)$ as the center and $\tau d$ as the radius. For the target $({\rho }'_{r},{{\theta }'_{r}})$ with an angle deviation from the transmitter satisfying $\tau d<{\tau }'d<\omega $, its average SNR is less than the threshold although it is within the beam divergence angle, and the target will not feedback the optical signal, resulting in the acquisition failure. (b) Archimedean spiral scanning. $U$ is field of uncertainty. The potential locations ${{\rho }_{1}}$ and ${{\rho }_{2}}$ of the receiver are within the central ring filled with red, they are acquired at the origin and on the red spiral, respectively. The blue section represents the $k=1$ ring. Whereas ${{\rho }_{3}}$ and ${{\rho }_{4}}$ are located within the gray-filled $k=2$ ring, they are acquired on the blue and the black spirals, respectively.}
	\label{fig:2}
\end{figure}

The scanning details are depicted in Fig. \ref{fig:2} (a). The optical signal is a Gaussian beam, whose divergence angle corresponding to ${1}/{{{e}^{2}}}$ intensity radius is $\omega \ge {{\omega }_{limit}}$, where ${{\omega }_{limit}}$ is the minimum divergence angle that the laser device can modulate. The distance between adjacent spiral arms is $d$, denoted as spiral pitch. The receiver may fall anywhere between adjacent spiral arms. The ideal scenario is on the spiral while the least favorable place would be on the midpoint of the adjacent spiral arms. To accommodate for this variability, $0\le \tau \le {1}/{2}$ is extracted as the coverage factor, representing the ratio of maximum acquisition deflection angle to spiral pitch that meets specified SNR level at the receiver for a certain transmitted power. In other words, the transmitted power can cover the circular range of radius $\tau d$, wherein the average SNR is greater than the threshold $\bar{Q}$.

Considering that the far-field propagation distance is $R$, the transmitter and receiver loss are $s_t$ and $s_r$, respectively, the diameter of the receiving aperture is $D_r$, the proportion of split beam for acquisition is $s_s$, and the angle deviation between the transmitter pointing and LoS is $\varphi $. It should be noted that the distance between the satellite and the ground exceeds 1000 kilometers, so the radius of beam that reaches the receiving plane with a divergence angle of 10 micro radians is beyond 10 meters. On the other hand, the receiver aperture size is typically in the range of tens of centimeters. Based on these considerations, it can be assumed that the beam incident on the receiver aperture is uniform. Then the $h_t$ is given by \citep{toyoshima2002optimum}:
\begin{eqnarray}
	{{h}_{t}}(\varphi )=\frac{2{{s}_{t}}{{s}_{r}}{{s}_{s}}}{\pi {{R}^{2}}}\frac{1}{{{\omega }^{2}}}\exp \left( -\frac{2{{\varphi }^{2}}}{{{\omega }^{2}}} \right)\cdot \pi {{\left( \frac{{{D}_{r}}}{2} \right)}^{2}}
	\label{eq:4}
\end{eqnarray}
where $\varphi $ is a random variable influenced by platform vibration. The expectation of $\varphi $ is $\tau d$, Generally, the variance of $\varphi $ is isotropic, i.e., ${{\sigma }_{x}}={{\sigma }_{y}}=\sigma \ne 0$, and the probability density function (PDF) of $\varphi $ is the Rice distribution \citep{rice1948statistical,friederichs2016vibration}:
\begin{eqnarray}
	{{f}_{\varphi }}(\varphi )=\frac{\varphi }{{{\sigma }^{2}}}\exp \left( -\frac{{{\varphi }^{2}}+{{\tau }^{2}d^2}}{2{{\sigma }^{2}}} \right){{I}_{0}}\left( \frac{\varphi \tau d}{{{\sigma }^{2}}} \right)
	\label{eq:5}
\end{eqnarray}
where ${{I}_{0}}(\cdot )$ is the zero-order modified Bessel function of the first kind. Then $E\left[ h_{t}^{2} \right]$ is calculated as:
\begin{eqnarray}
	E\left[ h_{t}^{2} \right]={{\left( \frac{{{s}_{t}}{{s}_{r}}{{s}_{s}}D_{r}^{2}}{2 {{R}^{2}}{{\omega }^{2}}} \right)}^{2}}{{E}_{\varphi }}\left[ \exp \left( -\frac{4{{\varphi }^{2}}}{{{\omega }^{2}}} \right) \right] ={{\left( \frac{{{s}_{t}}{{s}_{r}}{{s}_{s}}D_{r}^{2}}{2{{R}^{2}}} \right)}^{2}}\frac{1}{{{\omega }^{2}}\left( {{\omega }^{2}}+8{{\sigma }^{2}} \right)}\exp \left( -\frac{4{{\tau }^{2}}{{d}^{2}}}{{{\omega }^{2}}+8{{\sigma }^{2}}} \right)
	\label{eq:6}
\end{eqnarray}

The atmospheric turbulence is modeled by the Gamma-Gamma distribution in order to cover a wide range of turbulence conditions \citep{al2001mathematical}. Subsequently, the $E\left[ h_{c}^{2} \right]$ is obtained as \citep{wang2010moment}:
\begin{eqnarray}
	E\left[ h_{c}^{2} \right]=\frac{\left( \alpha +1 \right)\left( \beta +1 \right){{\gamma }^{2}}}{\alpha \beta }
	\label{eq:7}
\end{eqnarray}
where $\gamma $ is the scale parameter. $\alpha $ and $\beta $ are large-scale and small-scale effective numbers, respectively, which can directly be linked to the physical parameter Rytov variance \citep{prokevs2009modeling} and the detailed expressions are presented in \citet{andrews2005laser}.

Combining Eqs. (\ref{eq:3}), (\ref{eq:6}), and (\ref{eq:7}) yields the specific expression of $P_t$:
\begin{eqnarray}
	{{P}_{t}}=\frac{2{{R}^{2}}}{{{s}_{t}}{{s}_{r}}{{s}_{s}}\gamma D_{r}^{2}{{R}_{r}}}\sqrt{\frac{\bar{Q}{{N}_{0}}\alpha \beta }{\left( \alpha +1 \right)\left( \beta +1 \right)}}\omega \sqrt{{{\omega }^{2}}+8{{\sigma }^{2}}}\exp \left( \frac{2{{\tau }^{2}}{{d}^{2}}}{{{\omega }^{2}}+8{{\sigma }^{2}}} \right)
	\label{eq:9}
\end{eqnarray}

For the LEO-to-ground FSOC with a limited narrow window, the transmitter needs to scan at the maximum power to achieve acquisition in the shortest time, namely $P_t$ is a constant. Then we solve the relationship between $\tau$, $d$, and $\omega$:
\begin{eqnarray}
	\tau d={{g}_{B,\sigma }}\left( \omega  \right)=\sqrt{\frac{{{\omega }^{2}}+8{{\sigma }^{2}}}{2}\ln \left( \frac{B}{\omega \sqrt{{{\omega }^{2}}+8{{\sigma }^{2}}}} \right)}
	\label{eq:11}
\end{eqnarray}
where $B=\frac{{{P}_{t}}{{s}_{t}}{{s}_{r}}{{s}_{s}}\gamma D_{r}^{2}{{R}_{r}}}{2{{R}^{2}}}\sqrt{\frac{\left( \alpha +1 \right)\left( \beta +1 \right)}{\bar{Q}{{N}_{0}}\alpha \beta }}$. Meanwhile, we obtain the constraint on $\omega$ from Eq. (\ref{eq:11}):
\begin{eqnarray}
	\ln \left( \frac{B}{\omega \sqrt{{{\omega }^{2}}+8{{\sigma }^{2}}}} \right)>0\to \omega <\sqrt{\sqrt{{{B}^{2}}+16{{\sigma }^{4}}}-4{{\sigma }^{2}}}
	\label{eq:10}
\end{eqnarray}

\subsection{Acquisition time model} \label{sec:2.2}
The polar coordinate $(\rho, \theta)$ in Fig. \ref{fig:2} (a) represents the scanning trajectory on the spiral arm, where $\theta$ increments by $2\pi$ with each spiral turn. The Archimedean spiral is parameterized as \citep{steinhaus1999mathematical}:
\begin{eqnarray}
	\rho =\frac{d}{2\pi }\theta 
	\label{eq:12}
\end{eqnarray}

The position of the receiver, also known as the initial pointing error, follows a Gaussian distribution with zero mean, and the variances $({{\kappa }_{x}},{{\kappa }_{y}})$ are equal in both the horizontal and vertical directions, i.e. ${{\kappa }_{x}}={{\kappa }_{y}}=\kappa$. The corresponding polar coordinate is $({{\rho }_{r}},{{\theta }_{r}})$, where the polar angle ${{\theta }_{r}}$ adheres to a uniform distribution $U(0,2\pi )$, and the radial ${{\rho }_{r}}$ obeys the Rayleigh distribution as:
\begin{eqnarray}
	{{f}_{{{\rho }_{r}}}}({{\rho }_{r}})=\frac{{{\rho }_{r}}}{{{\kappa }^{2}}}\exp \left( -\frac{\rho _{r}^{2}}{2{{\kappa }^{2}}} \right)
	\label{eq:13}
\end{eqnarray}

In addition, we introduce the definition of a ring, which is the area enclosed by spirals with an increment of $\theta$ by $2\pi$, as depicted in Fig. \ref{fig:2} (b). Notably, the ring having $\theta \in [0,2\pi )$ is considered special because its inner degenerates to the origin and the distance between the outer spiral and the inner spiral (origin) is $\frac{d}{2\pi }\theta <d$, thus named the central ring. The others in the range of $\theta \in [2k\pi ,2k\pi +2\pi )$, $k=1,2,\ldots $, are specified as $k$ rings, where the distance between the outer and the inner spirals is $d$. For each $k$ ring, the inner spiral refers to the outer spiral of the previous ring, and correspondingly, the outer spiral denotes the inner spiral of the next ring. Consequently, by combining Eqs. (\ref{eq:12}), (\ref{eq:13}), and ${{\theta }_{r}}\sim U(0,2\pi )$, we obtain the PDF with respect to the coverage factor $\tau $ as:
\begin{eqnarray}
	\begin{aligned}
		& {{f}_{\tau }}(\tau )=\frac{1}{2\pi }\int{\left[ {{f}_{{{\rho }_{r}}}}\left( {{\rho }_{1}} \right){{\left| \frac{d{{\rho }_{1}}}{d\tau } \right|}_{{{\rho }_{1}}=\tau d}}+{{f}_{{{\rho }_{r}}}}\left( {{\rho }_{2}} \right){{\left| \frac{d{{\rho }_{2}}}{d\tau } \right|}_{{{\rho }_{2}}=(\frac{{{\theta }_{r}}}{2\pi }-\tau )d}} \right]d{{\theta }_{r}}} \\ 
		& \ \ \ \ \ \ \ \ +\ \frac{1}{2\pi }\sum\limits_{k=1}^{\infty }{\int{\left[ {{f}_{{{\rho }_{r}}}}\left( {{\rho }_{3}} \right){{\left| \frac{d{{\rho }_{3}}}{d\tau } \right|}_{{{\rho }_{3}}=(\frac{{{\theta }_{r}}}{2\pi }+k-1+\tau )d}}+{{f}_{{{\rho }_{r}}}}\left( {{\rho }_{4}} \right){{\left| \frac{d{{\rho }_{4}}}{d\tau } \right|}_{{{\rho }_{4}}=(\frac{{{\theta }_{r}}}{2\pi }+k-\tau )d}} \right]d{{\theta }_{r}}}} \\ 
	\end{aligned}
	\label{eq:14}
\end{eqnarray}
where the acquisition point of ${{\rho }_{1}}$ is the central ring origin, the position for ${{\rho }_{2}}$ is on the outer spiral of the central ring, whereas ${{\rho }_{3}}$ and ${{\rho }_{4}}$ are acquired on the inner and the outer spirals of the $k$ ring, as illustrated in Fig. \ref{fig:2} (b).  For $k$ rings, the distance from the receiver to the spiral where the acquisition point is located is $\tau d$, while the distance to the other spiral is $(1-\tau )d$. Since $0\le \tau \le \frac{1}{2}$, it always holds true that $\tau d\le (1-\tau )d$ for any $[0,2\pi]$. As for the central ring, the distance from the receiver to the acquisition spiral is also $\tau d$, but the distance to the other spiral is $\frac{d}{2\pi}{{\theta}_r}-\tau d$, which means that ${\theta}_r$ needs to satisfy the additional constraint $\tau d\le \frac{d}{2\pi}{{\theta}_r}-\tau d$ for a given $\tau$, i.e., $4\pi \tau \le {{\theta }_{r}}\le 2\pi $. Then Eq. (\ref{eq:14}) is integrated as:
\begin{eqnarray}
	\begin{aligned}
		& {{f}_{\tau }}(\tau )=\left( \frac{(1-2\tau )\tau {{d}^{2}}}{{{\kappa }^{2}}}+1 \right)\exp \left( -\frac{{{\tau }^{2}}{{d}^{2}}}{2{{\kappa }^{2}}} \right)-\exp \left( -\frac{{{(\tau -1)}^{2}}{{d}^{2}}}{2{{\kappa }^{2}}} \right) \\ 
		& \ \ \ \ \ \ \ \ +\sum\limits_{k=1}^{\infty }{\left[ \exp \left( -\frac{{{(k-1+\tau )}^{2}}{{d}^{2}}}{2{{\kappa }^{2}}} \right)-\exp \left( -\frac{{{(k+\tau )}^{2}}{{d}^{2}}}{2{{\kappa }^{2}}} \right) +\exp \left( -\frac{{{(k-\tau )}^{2}}{{d}^{2}}}{2{{\kappa }^{2}}} \right) -\exp \left( -\frac{{{(k+1-\tau )}^{2}}{{d}^{2}}}{2{{\kappa }^{2}}} \right) \right]} \\ 
		& \ \ \ \ \ \ \ =\left[ \frac{(1-2\tau )\tau {{d}^{2}}}{{{\kappa }^{2}}}+2 \right]\exp \left( -\frac{{{\tau }^{2}}{{d}^{2}}}{2{{\kappa }^{2}}} \right) \\ 
	\end{aligned}
	\label{eq:15}
\end{eqnarray}

When the closest distance between the spirals and the receiver is less than $\tau d$, the received average SNR is greater than the threshold, namely the transmitted power can cover the receiver given the SNR threshold, yielding a successful detection. Hence, the probability $P_{SNR}$ that the received average SNR over the threshold is the CDF of coverage factor:
\begin{eqnarray}
	{{P}_{SNR}}=P(\mathcal{T}\le \tau )=\int_{0}^{\tau }{{{f}_{\tau }}(\mathcal{T})d\mathcal{T}}=1+\left( 2\tau -1 \right)\exp \left( -\frac{{{\tau }^{2}}{{d}^{2}}}{2{{\kappa }^{2}}} \right)
	\label{eq:16}
\end{eqnarray}

Actually, $\kappa$ is milliradian magnitude, and $d$ is microradian magnitude, i.e., $\kappa \gg d$. Hence, $\tau$ approximately obeys uniform distribution $U(0,{1}/{2})$, and Eq. (\ref{eq:16}) is reduced as ${{P}_{SNR}}\approx 2\tau $.

Moreover, the signal is likely to be detected when LoS is within the field angle range of the receiver. The corresponding field detection probability $P_V$ is obtained as:
\begin{eqnarray}
	{{P}_{V}}=\int_{0}^{V}{{{f}_{{{\rho }_{r}}}}({{\rho }_{r}})d{{\rho }_{r}}}=1-\exp \left( -\frac{{{V}^{2}}}{2{{\kappa }^{2}}} \right)
	\label{eq:17}
\end{eqnarray}
where $V$ represents the half-width of the field angle of the photoelectric sensor at the receiver. Hence $0<{{P}_{V}}<1$ is independent of the scanning parameters of the transmitter and can be regarded as a constant. Then we define the feedback probability $P_R$ of the receiver as:
\begin{eqnarray}
	{{P}_{R}}={{P}_{V}}\cdot {{P}_{SNR}}=2{{P}_{V}}\tau <1
	\label{eq:18}
\end{eqnarray}

Furthermore, the acquisition is likely to be successful when the receiver $({{\rho }_{r}},{{\theta }_{r}})$ within the range of FOU, defined as $U$ and shown in Fig. \ref{fig:2} (b). The corresponding probability $P_U$ is expressed as \citep{li2011analytical}:
\begin{eqnarray}
	{{P}_{U}}=\int_{0}^{U}{{{f}_{{{\rho }_{r}}}}\left( {{\rho }_{r}} \right)d{{\rho }_{r}}}=1-\exp \left( -\frac{{{U}^{2}}}{2{{\kappa }^{2}}} \right)
	\label{eq:19}
\end{eqnarray}

Consequently, the probability of single-scan acquisition is:
\begin{eqnarray}
	{{P}_{S}}={{P}_{U}}\cdot {{P}_{R}}
	\label{eq:20}
\end{eqnarray}

As shown in Fig. \ref{fig:2} (b), the scanning usually adopts the Archimedean spiral to achieve an efficient search from high probability to low probability regions. We define $a={d}/{(2\pi )}$, and the length of the Archimedean spiral is given by \citet{steinhaus1999mathematical}:
\begin{eqnarray}
	L=\frac{1}{2}\left[ {{\rho }_{r}}\sqrt{1+{{\left( \frac{{{\rho }_{r}}}{a} \right)}^{2}}}+a\ln \left( \frac{{{\rho }_{r}}}{a}+\sqrt{1+{{\left( \frac{{{\rho }_{r}}}{a} \right)}^{2}}} \right) \right]
	\label{eq:21}
\end{eqnarray}

In general ${{\rho }_{r}}\gg a$ so that Eq. (\ref{eq:21}) is approximated by $L\approx {\rho _{r}^{2}}/{(2a)}$. Then the single-scan acquisition time ${t}_{S}$ is calculated with constant scanning speed $v$ as:
\begin{eqnarray}
	{{t}_{S}}=\frac{\pi \rho _{r}^{2}}{vd}
	\label{eq:22}
\end{eqnarray}

Combined with Eq. (\ref{eq:13}), the PDF of ${t}_{S}$ is obtained as:
\begin{eqnarray}
	{{f}_{{{t}_{S}}}}\left( {{t}_{S}} \right)=\frac{vd}{2\pi {{\kappa }^{2}}}\exp \left( -\frac{vd}{2\pi {{\kappa }^{2}}}{{t}_{S}} \right)
	\label{eq:23}
\end{eqnarray}

Subsequently, the time ${T}_{U}$ for scanning the complete FOU and the time expectation ${T}_{S}$ for the single-scan acquisition success are \citep{li2011analytical}:
\begin{eqnarray}
	{{T}_{U}}=\frac{\pi {{U}^{2}}}{vd}
	\label{eq:24}
\end{eqnarray}
\begin{eqnarray}
	{{T}_{S}}=\int_{0}^{{{T}_{U}}}{{{t}_{S}}\cdot {{f}_{{{t}_{S}}}}\left( {{t}_{S}} \right)d{{t}_{S}}}=\frac{2\pi {{\kappa }^{2}}}{vd}\left[ 1-\exp \left( -\frac{{{U}^{2}}}{2{{\kappa }^{2}}} \right)\left( 1+\frac{{{U}^{2}}}{2{{\kappa }^{2}}} \right) \right]
	\label{eq:25}
\end{eqnarray}

\section{Multi-scan mathematical model} \label{sec:3}
Acquisition success cannot be guaranteed with only once single-scan due to ${P}_{S}<1$. Therefore, the multi-scan, which is a series of repetitive scans over the same FOU, is often employed instead. In particular, when a single-scan proves failure, it is necessary to reinitialize the transmitter pointing based on ephemeris table and repeat the single-scan until a successful acquisition is achieved \citep{li2011analytical}. In this process, reset time ${{T}_{a}}$ for the APT to reinitialize the pointing to prepare for the next single-scan is strongly essential but ignored by previous analytical models. Therefore, when the acquisition is achieved in $n+1$ single-scan, the total scanning time ${{t}_{M}}$ is:
\begin{eqnarray}
	{{t}_{M}}=n\left( {{T}_{U}}+{{T}_{a}} \right)+{{t}_{S}}
	\label{eq:26}
\end{eqnarray}

The PDF of ${t}_{M}$ is:
\begin{eqnarray}
	{{f}_{{{t}_{M}}}}\left( {{t}_{M}} \right)={{\left( 1-{{P}_{S}} \right)}^{n}}{{P}_{R}}{{f}_{{{t}_{S}}}}\left( {{t}_{S}} \right)
	\label{eq:27}
\end{eqnarray}

Then we calculate the CDF of ${t}_{M}$ as:
\begin{eqnarray}
	\begin{aligned}
		& P\left( T\le {{t}_{M}} \right)=P\left( T\le n\left( {{T}_{U}}+{{T}_{a}} \right) \right)+P\left( n\left( {{T}_{U}}+{{T}_{a}} \right)<T\le {{t}_{M}} \right) \\ 
		& \ \ \ \ \ \ \ \ \ \ \ \ \ \ \ \ \ \ =\sum\limits_{k=0}^{n-1}{{{\left( 1-{{P}_{S}} \right)}^{k}}{{P}_{R}}\int_{0}^{{{T}_{U}}}{{{f}_{{{t}_{S}}}}\left( T \right)dT}}+{{(1-{{P}_{S}})}^{n}}{{P}_{R}}\int_{0}^{{{t}_{S}}}{{{f}_{{{t}_{S}}}}\left( T \right)dT} \\ 
		& \ \ \ \ \ \ \ \ \ \ \ \ \ \ \ \ \ \ =1-{{\left( 1-{{P}_{S}} \right)}^{n}}\left\{ 1-{{P}_{R}}\left[ 1-\exp \left( -\frac{vd}{2\pi {{\kappa }^{2}}}{{t}_{S}} \right) \right] \right\} \\ 
	\end{aligned}
	\label{eq:28}
\end{eqnarray}

There is $n\to \infty $ and ${{t}_{S}}\to {{T}_{U}}$ for ${{t}_{M}}\to \infty $. Then the acquisition probability of Eq. (\ref{eq:28}) becomes:
\begin{eqnarray}
	\underset{{{t}_{M}}\to \infty }{\mathop{\lim }}\,P\left( T\le {{t}_{M}} \right)=\underset{n\to \infty }{\mathop{\lim }}\,\left[ 1-{{\left( 1-{{P}_{S}} \right)}^{n+1}} \right]=1
	\label{eq:29}
\end{eqnarray}
which proves that the multi-scan can ensure acquisition success. The acquisition time expectation ${T}_{M}$ with multi-scan is calculated as:
\begin{eqnarray}
	{{T}_{M}}=\int_{0}^{\infty }{{{t}_{M}}\cdot {{f}_{{{t}_{M}}}}\left( {{t}_{M}} \right)d{{t}_{M}}} =\frac{{{T}_{S}}}{{{P}_{U}}}+\left( \frac{1}{{{P}_{S}}}-1 \right)\left( {{T}_{U}}+{{T}_{a}} \right) =\frac{2\pi {{\kappa }^{2}}}{vd}\left[ \frac{{{e}^{\eta }}\eta \left( 1-{{P}_{R}} \right)}{\left( {{e}^{\eta }}-1 \right){{P}_{R}}}+1 \right]+\frac{{{T}_{a}}\left[ {{e}^{\eta }}\left( 1-{{P}_{R}} \right)+{{P}_{R}} \right]}{\left( {{e}^{\eta }}-1 \right){{P}_{R}}}
	\label{eq:30}
\end{eqnarray}
where $\eta ={{{U}^{2}}}/{(2{{\kappa }^{2}})}>0$. Given that $\omega$, $d$, and $\tau$ are coupled according to Eq. (\ref{eq:11}), where any two known terms can solve the remaining one theoretically. However, $\omega$ cannot be uniquely determined by $\tau$ and $d$. Since $d$ only appears once in Eq. (\ref{eq:30}), we replace $d$ to facilitate subsequent optimization analysis for the goal of minimizing ${T}_{M}$.

\subsection{Spiral pitch optimization}
When ${{{g}_{B}}\left( \omega  \right)}/{d}>{1}/{2}$, there is $\tau \equiv {1}/{2}$ with ${{P}_{R}}={{P}_{V}}$ in Eq. (\ref{eq:30}), where ${{T}_{M}}$ is not related to $\tau $ and it decreases with $d$. The minimum is taken at $d=2{{g}_{B,\sigma }}\left( \omega  \right)$.

While ${{{g}_{B}}\left( \omega  \right)}/{d}\le {1}/{2}$, the derivative of ${T}_{M}$ with respect to $\tau $ by replacing $d={{{g}_{B,\sigma }}\left( \omega  \right)}/{\tau }$ in Eq. (\ref{eq:30}) is:
\begin{eqnarray}
	\frac{\partial {{T}_{M}}}{\partial \tau }=\frac{2\pi {{\kappa }^{2}}\left( {{e}^{\eta }}-{{e}^{\eta }}\eta -1 \right)}{v\left( {{e}^{\eta }}-1 \right)\cdot {{g}_{B,\sigma }}\left( \omega  \right)}-\frac{2{{e}^{\eta }{{P}_{V}}}{{T}_{a}}}{\left( {{e}^{\eta }}-1 \right)P_{R}^{2}}
	\label{eq:31}
\end{eqnarray}
where ${T}_{M}$ decreases monotonically with $\tau $ due to $({{e}^{\eta }}-{{e}^{\eta }}\eta -1)$ constantly less than zero. The minimum ${T}_{M}$ is taken at ${{\tau }_{opt}}={1}/{2}$. Consequently, the optimum spiral pitch ${{d}_{opt}}$ is:
\begin{eqnarray}
	{{d}_{opt}}=2{{g}_{B,\sigma }}\left( \omega  \right)=\sqrt{2\left( {{\omega }^{2}}+8{{\sigma }^{2}} \right)\ln \left( \frac{B}{\omega \sqrt{{{\omega }^{2}}+8{{\sigma }^{2}}}} \right)}
	\label{eq:32}
\end{eqnarray}

\subsection{Beam divergence angle optimization}
By the same token, ${{T}_{M}}$ is independent of $\omega $ when ${{{g}_{B}}\left( \omega  \right)}/{d}>{1}/{2}$, thus ${T}_{M}$ does not change with $\omega $ in this case.

While ${{{g}_{B}}\left( \omega  \right)}/{d}\le {1}/{2}$, the derivative of ${T}_{M}$ with respect to $\omega $ by replacing $d={{{g}_{B,\sigma }}\left( \omega  \right)}/{\tau }$ is:
\begin{eqnarray}
	\frac{\partial {{T}_{M}}}{\partial \omega }=\frac{\partial {{T}_{M}}}{\partial {{g}_{B,\sigma }}} \frac{\partial {{g}_{B,\sigma }}}{\partial \omega }=-\frac{\pi {{\kappa }^{2}}\left[ {{e}^{\eta }}\eta \left( 1-{{P}_{R}} \right)+{{P}_{R}}\left( {{e}^{\eta }}-1 \right) \right]}{v\left( {{e}^{\eta }}-1 \right){{P}_{V}}\cdot g_{B,\sigma }^{2}\left( \omega  \right)} \times \frac{{{\omega }^{2}}\ln \left( \frac{B}{\omega \sqrt{{{\omega }^{2}}+8{{\sigma }^{2}}}} \right)-{{\omega }^{2}}-4{{\sigma }^{2}}}{2\omega \cdot {{g}_{B,\sigma }}\left( \omega  \right)}
	\label{eq:33}
\end{eqnarray}

Let ${\partial {{T}_{M}}}/{\partial \omega }=0$, i.e., ${\partial {{g}_{B,\sigma }}}/{\partial \omega }=0$, and obtain:
\begin{eqnarray}
	B={{B}_{\sigma }}\left( \omega  \right)=\omega \sqrt{{{\omega }^{2}}+8{{\sigma }^{2}}}\exp \left( 1+\frac{4{{\sigma }^{2}}}{{{\omega }^{2}}} \right)
	\label{eq:34}
\end{eqnarray}

The derivative of ${{B}_{\sigma }}\left( \omega  \right)$ concerning $\omega $ is:
\begin{eqnarray}
	\frac{\partial {{B}_{\sigma }}}{\partial \omega }=\frac{2\left( {{\omega }^{4}}-32{{\sigma }^{4}} \right)}{{{\omega }^{2}}\sqrt{{{\omega }^{2}}+8{{\sigma }^{2}}}}\exp \left( 1+\frac{4{{\sigma }^{2}}}{{{\omega }^{2}}} \right)
	\label{eq:35}
\end{eqnarray}
where ${{B}_{\sigma }}$ decreases for $\omega <{{2}^{5/4}}\sigma $ and increases for $\omega >{{2}^{5/4}}\sigma $. The minimum ${{B}_{\sigma }}$ is reached at $\omega ={{2}^{5/4}}\sigma $, namely $B_{\sigma }^{\min }={{B}_{\sigma }}\left( {{2}^{{5}/{4}}}\sigma \right)$.

Thereby, when $B\le B_{\sigma }^{\min }$, there is ${\partial {{g}_{B,\sigma }}}/{\partial \omega }<0$ and ${\partial {{T}_{M}}}/{\partial \omega }>0$, ${T}_{M}$ increases monotonically with $\omega $ in this case, and the minimum value is taken at ${{\omega }_{limit}}$.

When $B>B_{\sigma }^{\min }$, ${T}_{M}$ has two extreme points ${{B}_{\sigma }}\left( {{\omega }_{top}} \right)={{B}_{\sigma }}\left( {{\omega }_{btm}} \right)=B$, where ${{\omega }_{top}}$ is the maximum point in $\left( 0,{{2}^{5/4}}\sigma  \right]$, and ${{\omega }_{btm}}$ is the minimum in $\left( {{2}^{5/4}}\sigma ,+\infty  \right)$, both are within the range of Eq. (\ref{eq:10}). Obviously, there must be a point ${{\omega }_{eq}}\in \left( 0,{{2}^{5/4}}\sigma  \right]$ that satisfies ${{T}_{M}}\left( {{\omega }_{eq}} \right)={{T}_{M}}\left( {{\omega }_{btm}} \right)$. Consequently, the minimum ${T}_{M}$ is obtained at ${{\omega }_{btm}}$ for ${{\omega }_{eq}}\le {{\omega }_{limit}}\le {{\omega }_{btm}}$ or at ${{\omega }_{limit}}$ for ${{\omega }_{limit}}<{{\omega }_{eq}}$ or ${{\omega }_{limit}}>{{\omega }_{btm}}$ as:
\begin{eqnarray}
	T_{M}^{\min }=\left\{ \begin{aligned}
		& {{T}_{M}}({{\omega }_{btm}}),\ {{\omega }_{eq}}\le \ {{\omega }_{limit}}\le \ {{\omega }_{btm}} \\ 
		& {{T}_{M}}({{\omega }_{limit}}),\ \text{else} \\ 
	\end{aligned} \right.
	\label{eq:36}
\end{eqnarray}

Since the analytical expression of ${\omega }_{eq}$ is unsolvable, applying Eq. (\ref{eq:36}) directly becomes challenging. However, we can take advantage of the known quantities $B$ and ${{\omega }_{limit}}$ to determine the optimum beam divergence angle by indirect derivation and comparison.

As deduced before, ${{B}_{\sigma }}\left( {{\omega }_{limit}} \right)$ monotonically increases when ${{\omega }_{limit}}\ge {{2}^{5/4}}\sigma $. If ${{B}_{\sigma }}\left( {{\omega }_{btm}} \right)=B<{{B}_{\sigma }}\left( {{\omega }_{limit}} \right)$, there is ${{\omega }_{btm}}<{{\omega }_{limit}}$ and the minimum ${T}_{M}$ is obtained at ${{\omega }_{limit}}$, otherwise at ${{\omega }_{btm}}$.

It can be found from Eq. (\ref{eq:33}) that the trends of ${T}_{M}$ and ${{g}_{B,\sigma }}$ with $\omega $ are opposite. Thereby the minimum of ${{g}_{B,\sigma }}$ for $\omega \in \left( 0,{{2}^{5/4}}\sigma  \right]$ is taken at ${{\omega }_{top}}$, and the corresponding minimum is calculated by substituting $B={{B}_{\sigma }}\left( {{\omega }_{top}} \right)$ into ${{g}_{B,\sigma }}\left( {{\omega }_{top}} \right)$ as:
\begin{eqnarray}
	{{g}_{B,\sigma }}\left( \left. \omega  \right|\omega \le {{2}^{5/4}}\sigma  \right)\ge {{g}_{B,\sigma }}\left( {{\omega }_{top}} \right)=\sqrt{\frac{\left( \omega _{top}^{2}+8{{\sigma }^{2}} \right)\left( \omega _{top}^{2}+4{{\sigma }^{2}} \right)}{2\omega _{top}^{2}}}\ge \left( 2+\sqrt{2} \right)\sigma 
	\label{eq:37}
\end{eqnarray}

If ${{\omega }_{eq}}$ is known, ${{\omega }_{btm}}$ can be solved by ${{g}_{B,\sigma }}\left( {{\omega }_{eq}} \right)={{g}_{B,\sigma }}\left( {{\omega }_{btm}} \right)$ as:
\begin{eqnarray}
	{{\omega }_{btm}}=W\left( A \right)=\sqrt{{{A}^{2}}-6{{\sigma }^{2}}+\sqrt{{{A}^{4}}-12{{A}^{2}}+4{{\sigma }^{4}}}}>{{2}^{5/4}}\sigma 
	\label{eq:38}
\end{eqnarray}
where $A={{g}_{B,\sigma }}\left( {{\omega }_{eq}} \right)$ with constraint $A\ge \left( 2+\sqrt{2} \right)\sigma $, which is exactly consistent with Eq. (\ref{eq:37}), thus ${{\omega }_{btm}}$ certainly exist and increases monotonically with $A$. Additionally, we obtain ${{\omega }'_{btm}}$ by substituting $A'={{g}_{B,\sigma }}\left( {{\omega }_{limit}} \right)$ into Eq. (\ref{eq:38}) for ${{\omega }_{limit}}<{{2}^{5/4}}\sigma $. If $B={{B}_{\sigma }}\left( {{\omega }_{btm}} \right)<{{B}_{\sigma }}\left( {{\omega }'_{btm}} \right)$, there is ${\omega }_{btm}<{{\omega }'_{btm}}$, further ${{g}_{B,\sigma }}\left( {{\omega }_{eq}} \right)=A<A'={{g}_{B,\sigma }}\left( {{\omega }_{limit}} \right)$, thus ${{T}_{M}}\left( {{\omega }_{limit}} \right)<{T}_{M}\left( {{\omega }_{eq}} \right)={{T}_{M}}\left( {{\omega }_{btm}} \right)$, the minimum ${T}_{M}$ is obtained at ${{\omega }_{limit}}$. Otherwise at ${{\omega }_{btm}}$. Consequently, the optimum beam divergence angle ${{\omega }_{opt}}$ is:
\begin{eqnarray}
	{{\omega }_{opt}}=\left\{ \begin{aligned}
		& {{\omega }_{limit}},\begin{matrix}
			B<{{B}_{\sigma }}\left( W\left( {{g}_{B,\sigma }}({{\omega }_{limit}}) \right) \right)\ \text{with}\ {{\omega }_{limit}}<{{2}^{5/4}}\sigma   \\
			B<{{B}_{\sigma }}\left( {{\omega }_{limit}} \right)\ \text{with}\ {{\omega }_{limit}}\ge {{2}^{5/4}}\sigma   \\
		\end{matrix} \ \ \text{or} \\ 
		& {{\omega }_{btm}},\ \text{else} \\ 
	\end{aligned} \right.
	\label{eq:39}
\end{eqnarray}

When $\omega _{btm}^{2}\gg 4{{\sigma }^{2}}$, Eq. (\ref{eq:34}) is approximated as:
\begin{eqnarray}
	\begin{aligned}
		& \frac{B}{e}=\omega \sqrt{{{\omega }^{2}}+8{{\sigma }^{2}}}\cdot \exp \left( \frac{4{{\sigma }^{2}}}{{{\omega }^{2}}} \right)\approx \omega \sqrt{{{\omega }^{2}}+8{{\sigma }^{2}}}\left( 1+\frac{4{{\sigma }^{2}}}{{{\omega }^{2}}} \right) \\ 
		& \ \ \ =\left( {{\omega }^{2}}+4{{\sigma }^{2}} \right)\sqrt{1+\frac{8{{\sigma }^{2}}}{{{\omega }^{2}}}}\approx \left( {{\omega }^{2}}+4{{\sigma }^{2}} \right)\left( 1+\frac{4{{\sigma }^{2}}}{{{\omega }^{2}}} \right)={{\left( \omega +\frac{4{{\sigma }^{2}}}{\omega } \right)}^{2}} \\ 
	\end{aligned}
	\label{eq:40}
\end{eqnarray}

Therefore, the approximate analytical ${\omega }_{btm}$ can be solved, whose the difference from the numerical solutions is within $0.1\%$ in the case of $B \ge 2B_{\sigma }^{\min }$.

However, the value of ${{\omega }_{btm}}$ decreases as $B$ get smaller than $2B_{\sigma }^{\min }$ so that the approximate error becomes larger. We adopt polynomials to fit the numerical solutions of Eq. (\ref{eq:34}), where the goodness of fit (GoF) is utilized as an index to evaluate the fitting accuracy. The variable is $x={B}/{B_{\sigma }^{\min }}$ and fit in $\left[ 1,2 \right)$ with $GoF=0.999$. The piecewise ${\omega }_{btm}$ is expressed as:
\begin{eqnarray}
	{{\omega }_{btm}}=\left\{ \begin{aligned}
		& \left( 1.5087{{x}^{3}}-7.9617{{x}^{2}}+15.913x-6.8278 \right)\sigma ,\ \ 1\le x<2 \\ 
		& \frac{\sqrt{B_{\sigma }^{\min }x}+\sqrt{B_{\sigma }^{\min }x-16e{{\sigma }^{2}}}}{2\sqrt{e}},\ \ \ \ \ \ \ \ \ \ \ \ \ \ \ \ \ \ \ \ \ \ \ \ \ \ \ \ \ \ 2\le x \\ 
	\end{aligned} \right.
	\label{eq:41}
\end{eqnarray}

\subsection{FOU optimization}
The derivative of ${T}_{M}$ with respect to $\eta $ is:
\begin{eqnarray}
	\frac{\partial {{T}_{M}}}{\partial \eta }=\frac{2\pi {{\kappa }^{2}}{{e}^{\eta }}\left( 1-{{P}_{R}} \right)\left( {{e}^{\eta }}-\eta -1-{{{\hat{T}}}_{a}} \right)}{vd{{\left( {{e}^{\eta }}-1 \right)}^{2}}{{P}_{R}}}
	\label{eq:42}
\end{eqnarray}
where ${{\hat{T}}_{a}}=\frac{vd\cdot {{T}_{a}}}{2\pi {{\kappa }^{2}}\left( 1-{{P}_{R}} \right)}$, the minimum ${T}_{M}$ is taken at ${\partial {{T}_{M}}}/{\partial \eta }=0$ as:
\begin{eqnarray}
	{{e}^{\eta }}-\eta -1-{{\hat{T}}_{a}}=0
	\label{eq:43}
\end{eqnarray}

However, the above equation has no analytical solution. When $\eta $ is large, there is approximate $\eta \approx \ln ({{\hat{T}}_{a}})$, which can be employed as the variable to perform polynomial fitting with numerical solutions and effectively reduce the order. In general, $v$ and $\kappa $ are of the same order of magnitude, which is three orders larger than $d$. ${T}_{a}$ and $(1-{{P}_{R}})$ are the level of ${10}^{1}$ and ${10}^{-2}$, respectively. Therefore, ${{\hat{T}}_{a}}$ is about ${{10}^{-1}}\sim{{10}^{0}}$ magnitude order. Without loss of generality, we perform piecewise fitting in the interval $\left[ 0.01,10 \right]$ with $GoF=0.999$. Consequently, the fitting polynomials for $x=\ln ({{\hat{T}}_{a}})$ is:
\begin{eqnarray}
	{{\eta }_{opt}}=\left\{ \begin{aligned}
		& 0.02824{{x}^{2}}+0.3137x+0.9873,\ \ 0.01\le {{{\hat{T}}}_{a}}<0.1 \\ 
		& 0.06114{{x}^{2}}+0.4549x+1.1445,\ \ \ \ 0.1\le {{{\hat{T}}}_{a}}\le 1 \\ 
		& 0.07171{{x}^{2}}+0.4725x+1.1441,\ \ \ \ \ \ 1<{{{\hat{T}}}_{a}}\le 10 \\ 
	\end{aligned} \right.
	\label{eq:44}
\end{eqnarray}

Then the optimum FOU is:
\begin{eqnarray}
	{{U}_{opt}}=\kappa \sqrt{2{{\eta }_{opt}}}
	\label{eq:45}
\end{eqnarray}

Further bring Eq. (\ref{eq:43}) into (\ref{eq:30}), the expression of minimum ${T}_{M}$ is obtained as:
\begin{eqnarray}
	T_{M}^{\min }=\frac{1}{{{P}_{R}}}\cdot \frac{2\pi {{\kappa }^{2}}}{vd}+\left( \frac{1}{{{P}_{R}}}-1 \right)\left( \frac{\pi U_{opt}^{2}}{vd}+{{T}_{a}} \right)
	\label{eq:46}
\end{eqnarray}
which is similar to the form of ${{T}_{M}}=\frac{{{T}_{S}}}{{{P}_{U}}}+\left( \frac{1}{{{P}_{S}}}-1 \right)\left( {{T}_{U}}+{{T}_{a}} \right)$ in Eq. (\ref{eq:30}). Given that ${2\pi {{\kappa }^{2}}}/{(vd)}=\int_{0}^{\infty }{{{t}_{S}}\cdot {{f}_{{{t}_{S}}}}\left( {{t}_{S}} \right)d{{t}_{S}}}$, we can equivalent the optimization conclusion of FOU to a multi-scan scenario where the receiver feedback probability is one, the single-scan acquisition probability is ${P}_{R}$, and the scanning range is at least ${U}_{opt}$.

\subsection{Platform vibration influence} \label{sec:3.4}
It can be found in Eq. (\ref{eq:39}) that the optimum beam divergence angle changes with the vibration standard deviation. Hence, the analysis of the vibration influence on the multi-scan acquisition time is significant. Through the principle of composite derivation, we get:
\begin{eqnarray}
	\frac{\partial {{T}_{M}}}{\partial \sigma }=\frac{\partial {{T}_{M}}}{\partial {{g}_{B,\sigma }}}\left( \frac{\partial {{g}_{B,\sigma }}}{\partial \sigma }+\frac{\partial {{g}_{B,\sigma }}}{\partial \omega }\frac{\partial \omega }{\partial \sigma } \right) =-\frac{\pi {{\kappa }^{2}}\left[ {{e}^{\eta }}\eta \left( 1-{{P}_{R}} \right)+{{P}_{R}}\left( {{e}^{\eta }}-1 \right) \right]}{v\left( {{e}^{\eta }}-1 \right){{P}_{V}}\cdot g_{B,\sigma }^{2}\left( \omega  \right)}\times \left[ \frac{4\ln \left( \frac{B}{\omega \sqrt{{{\omega }^{2}}+8{{\sigma }^{2}}}} \right)-2}{{{g}_{B,\sigma }}\left( \omega  \right)}+\frac{\partial {{g}_{B,\sigma }}}{\partial \omega }\frac{\partial \omega }{\partial \sigma } \right]
	\label{eq:47}
\end{eqnarray}

When ${{\omega }_{opt}}={{\omega }_{limit}}$, the divergence angle is constant thus ${\partial \omega }/{\partial \sigma }=0$. ${T}_{M}$ increases monotonically with $\sigma $ for ${{\omega }_{limit}}>{{B}^{{1}/{2}}}{{e}^{{-1}/{4}}}$. While for ${{\omega }_{limit}}\le {{B}^{{1}/{2}}}{{e}^{{-1}/{4}}}$, the optimum is taken at ${\partial {{T}_{M}}}/{\partial \sigma }=0$, i.e., ${{\sigma }_{opt}}={\sqrt{{{B}^{2}}-e{{\omega }^{4}}}}/{\left( 2\sqrt{2e}\omega  \right)}$. Therefore, the corresponding ${{\omega }_{\sigma ,limit}}$ is solved with the known standard deviation $\sigma $ of platform vibration as:
\begin{eqnarray}
	{{\omega }_{\sigma ,limit}}=\sqrt{\sqrt{{{{B}^{2}}}/{e}+16{{\sigma }^{4}}}-4{{\sigma }^{2}}}
	\label{eq:48}
\end{eqnarray}
which is in the range of Eq.(\ref{eq:10}). We design ${\omega }_{limit}$ by referring to Eq. (\ref{eq:48}), whose rationality is reflected in that the change of vibration intensity leads to an increase in acquisition time.

When ${{\omega }_{opt}}={{\omega }_{btm}}$, there is ${\partial {{g}_{B,\sigma }}}/{\partial \omega }=0$ and $4\ln \left( \frac{B}{\omega \sqrt{{{\omega }^{2}}+8{{\sigma }^{2}}}} \right)-2=2+\frac{16{{\sigma }^{2}}}{{{\omega }^{2}}}>0$ according to Eq. (\ref{eq:33}), further ${T}_{M}$ decreases with $\sigma $.

\section{Discussion of the results} \label{sec:4}
\begin{figure}[t!]
	\centering\includegraphics[width=0.65\columnwidth]{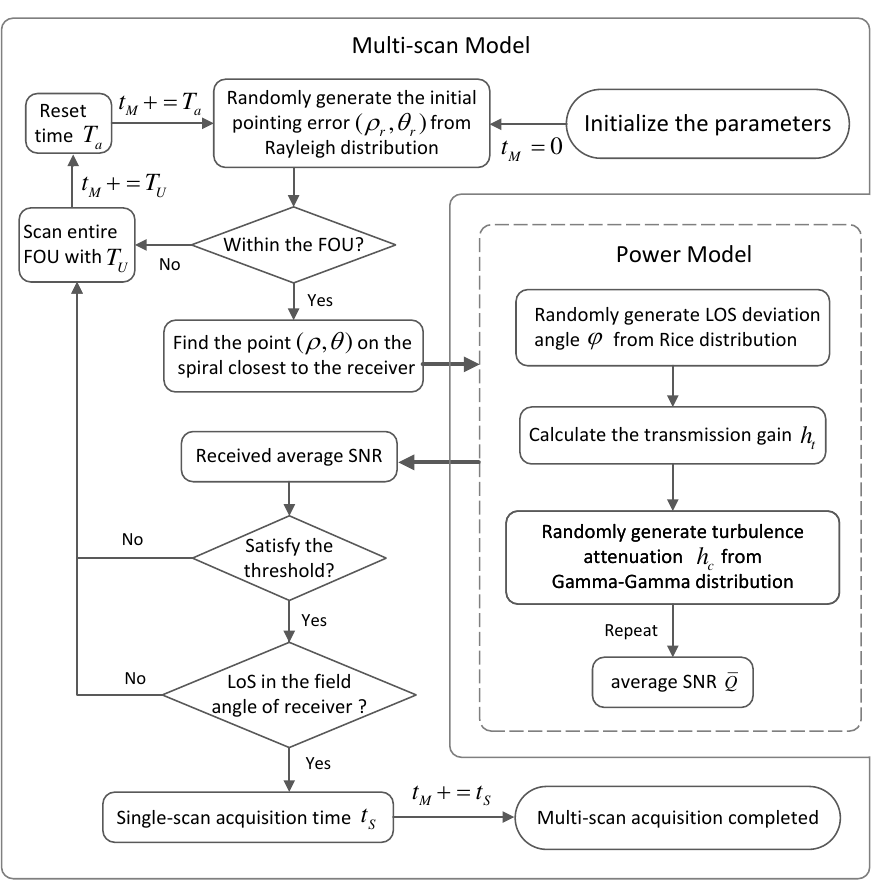}
	\caption{Simulation process of Monto Carlo}
	\label{fig:3}
\end{figure}

The numerical results are presented as an illustration of the above derivations. The MC simulation results are also performed to verify the obtained analytical expressions. The simulation process is illustrated in Fig. \ref{fig:3}, and the corresponding parameters are listed in Table \ref{tab:1}.

\begin{table*}[htbp!]
	\centering
	\caption{Simulation parameters}
	\label{tab:1}
	\begin{tabular}{@{}lll@{}}
		\toprule
		Parameters                        & Value & Unit/Remark   \\ \midrule
		Link distance $R$                    & 1200 & $km$     \\
		Transmitter loss $s_t$                  & 0.92  & --      \\
		Receiver loss $s_r$                     & 0.92  & --     \\
		Receiver aperture diameter $D_r$            & 30 & $cm$    \\
		Proportion of split beam $s_s$            & 0.1 & --    \\
		Photoelectric response efficiency $R_r$  & 0.88   & --      \\
		Std. of noise current ${\sigma }_n$             & 9   & $nA$ \\
		Receiver average SNR threshold $\bar{Q}$   & 20    & $dB$     \\
		Frequency of platform vibration $F_V$		& 100		& $Hz$	\\
		Std. of platform vibration $\sigma $       & 4     & $\mu rad$  \\
		Std. of initial LOS error $\kappa$         & 1  & $mrad$       \\
		Scanning speed $v$                    & 0.4  & $mrad/s$       \\
		Transmitter power ${P}_{t}$           & 90  & $mW$       \\
		Reset time ${T}_{a}$ 	 & 10  & $Sec.$       \\
		Field detection probability ${P}_{V}$  & 95\%   & --        \\ \midrule
		\multirow{5}{*}{Turbulence parameters $(\gamma ,\alpha ,\beta )$} &	$(0.90,21.6,19.8)$ &	$Turb.1$ very weak level		\\
		& $(0.58,8.43,6.92)$ & $Turb.2$ weak level \\
		& $(0.36,4.03,1.54)$ &  $Turb.3$ medium level	\\
		& $(0.27,4.58,1.24)$ &  $Turb.4$ strong level 	\\
		& $(0.21,6.07,1.08)$ &  $Turb.5$ very strong level	\\ \bottomrule
	\end{tabular}
\end{table*}

\begin{figure}[htbp!]
	\centering\includegraphics[width=0.5\columnwidth]{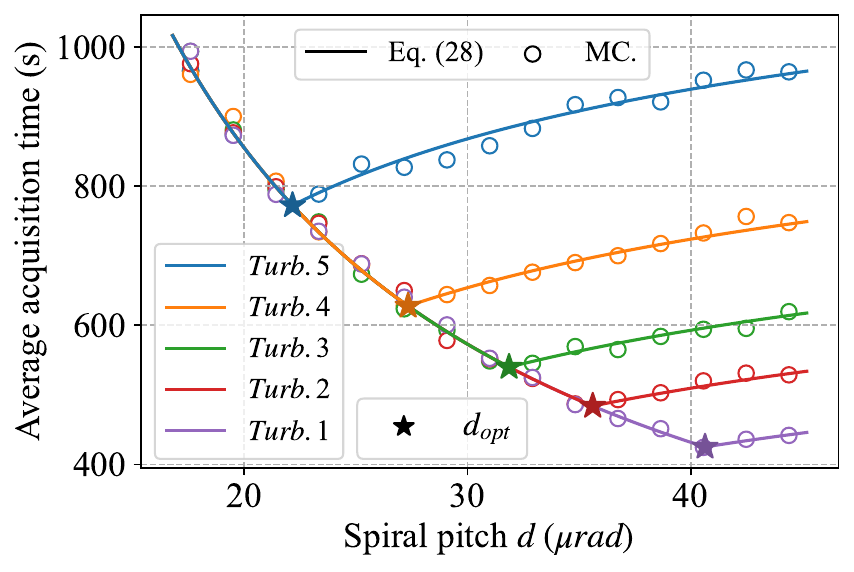}
	\caption{The variation of the multi-scan acquisition time with the spiral pitch under different turbulences.}
	\label{fig:4}
\end{figure}
Fig. \ref{fig:4} depicts the variation of the multi-scan acquisition time with the spiral pitch under different turbulences, where ${\omega }=20\mu rad$ and $U=1.3mrad$. The simulation process is illustrated as the multi-scan model in Fig. \ref{fig:3}. The theoretical acquisition time and the optimum spiral pitch can be calculated according to Eqs. (\ref{eq:30}) and (\ref{eq:32}), respectively. It should be noted that ${{{g}_{B,\sigma }}(\omega )}/{d}$ exceeds the upper limit of $\tau \in \left[ 0,{1}/{2} \right]$ for $d<{{d}_{opt}}$, thus the corresponding coverage factor is identified as ${1}/{2}$. It can be found that the theoretical results are in good agreement with the corresponding MC results, which demonstrate the multi-scan model. Moreover, the acquisition time increases with the spiral pitch when $d\ge {{d}_{opt}}$, which is because the decrease of the corresponding coverage factor results in the decrease of single-scan acquisition probability. While $d<{{d}_{opt}}$, there is ${{P}_{R}}={{P}_{V}}$, and the ${{T}_{M}}$ is positively correlated with ${2\pi {{\kappa }^{2}}}/{(vd)}$, hence ${{T}_{M}}$ decreases with $d$. Actually, this part has redundancy in the case of ${{P}_{SNR}}=1$ so that the acquisition time is equal under different turbulences. In addition, the smaller the spiral pitch, the more redundancy and the longer the time. Furthermore, as the turbulence increases, ${{g}_{B,\sigma }}(\omega )$ decreases, thus the ${{d}_{opt}}$ decreases gradually. Meanwhile, the $\tau $ decreases as ${{g}_{B,\sigma }}(\omega )$ decreases under the same $d$, resulting in the corresponding acquisition time increases. Consequently, the optimization conclusion of the spiral pitch is verified.

\begin{figure}[htbp!]
	\centering\includegraphics[width=0.5\columnwidth]{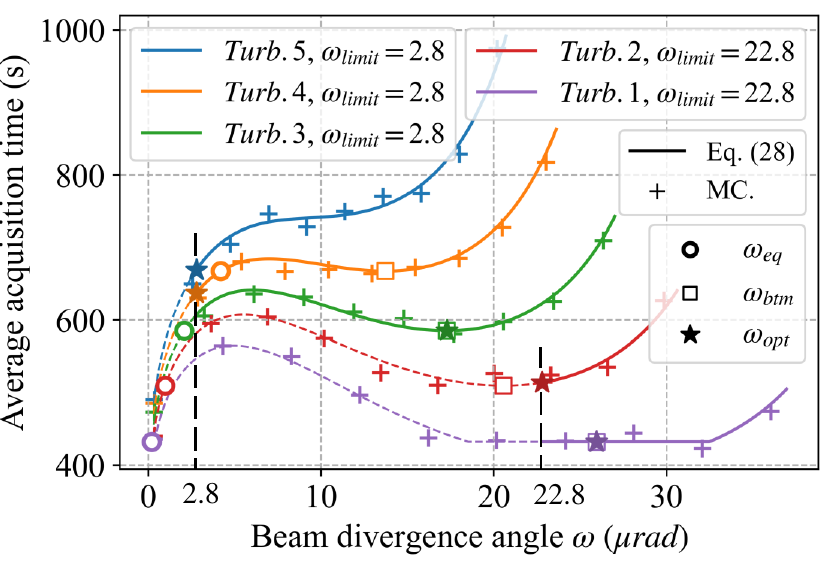}
	\caption{The variation of the multi-scan acquisition time with the beam divergence angle under different turbulences. The dashed lines represent the part where $\omega <{{\omega }_{limit}}$.}
	\label{fig:5}
\end{figure}
Fig. \ref{fig:5} presents the variation of the multi-scan acquisition time with the beam divergence angle under different turbulences, where $d=40\mu rad$ and $U=1.3mrad$. The theoretical optimum divergence angle is calculated from Eq. (\ref{eq:39}). Analogously, the numerical results are an excellent match with the corresponding MC results. For level $Turb.5$, we get $B=45.7{{\sigma }^{2}}<48.5{{\sigma }^{2}}=B_{\sigma }^{\min }$ so that the acquisition time increases monotonically with the divergence angle, and the optimum is obtained at ${{\omega }_{limit}}$. When turbulence level is $Turb.4$, although $B>B_{\sigma }^{\min }$, there still ${{\omega }_{opt}}={{\omega }_{limit}}$ due to ${{\omega }_{limit}}<{{\omega }_{eq}}$ with $B=58.3{{\sigma }^{2}}<{{B}_{\sigma }}\left( W\left( {{g}_{58.3{{\sigma }^{2}},\sigma }}(0.7\sigma ) \right) \right)=64{{\sigma }^{2}}$. As the turbulence weakens, $B$ increases and ${{\omega }_{eq}}$ decreases gradually. Then we obtain ${{\omega }_{eq}}<{{\omega }_{limit}}$ with $B=75.1{{\sigma }^{2}}>{{B}_{\sigma }}\left( W\left( {{g}_{75.1{{\sigma }^{2}},\sigma }}(0.7\sigma ) \right) \right)=70{{\sigma }^{2}}$ at $Turb.3$, where the optimum divergence angle is taken at ${{\omega }_{btm}}$. While $Turb.2$ with $B=95.5{{\sigma }^{2}}<{{B}_{\sigma }}\left( 5.7\sigma  \right)=111.5{{\sigma }^{2}}$, ${{\omega }_{btm}}<{{\omega }_{limit}}$ is not within the feasible range thereby ${{\omega }_{opt}}={{\omega }_{limit}}$. As the turbulence continues to weaken, $B$ and ${{\omega }_{btm}}$ gradually increase. We get ${{\omega }_{limit}}<{{\omega }_{btm}}$ with $B=137.3{{\sigma }^{2}}>111.5{{\sigma }^{2}}$ at $Turb.1$, where the optimum divergence angle is taken at ${{\omega }_{btm}}$. Moreover, the acquisition time does not change with $\omega $ between $18.8\mu rad$ and $32\mu rad$, which is because ${{{g}_{B,\sigma }}\left( \omega  \right)}/{d}>{1}/{2}$ and single-scan acquisition probability remains constant. Consequently, the optimization conclusion of the beam divergence angle is verified.

\begin{figure}[htbp!]
	\centering\includegraphics[width=0.5\columnwidth]{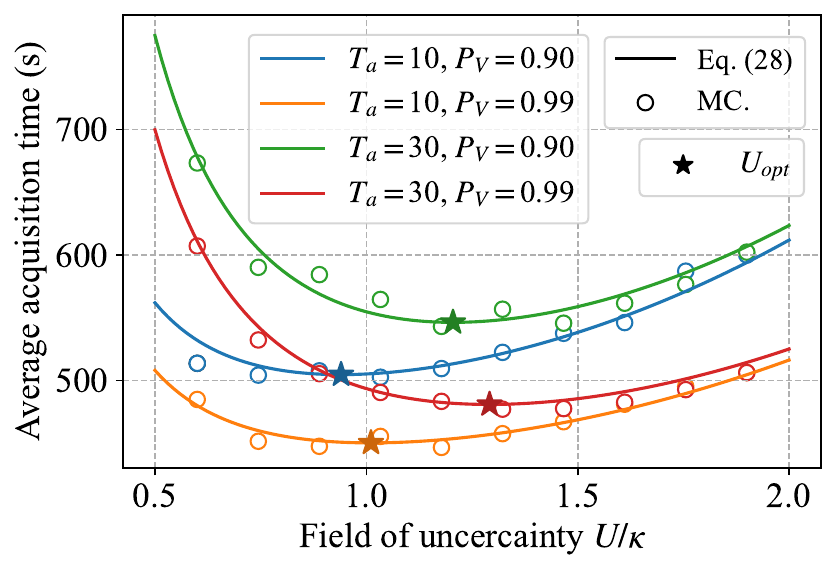}
	\caption{The variation of the multi-scan acquisition time with the FOU under different combinations of reset time and field detection probability at turbulence level $Turb.3$.}
	\label{fig:6}
\end{figure}
Fig. \ref{fig:6} shows the variation of the multi-scan acquisition time with the FOU under different combinations of reset time and field detection probability, where ${\omega }=20\mu rad$, $d=48\mu rad$, and turbulence level is $Turb.3$. The theoretical optimum FOU is fitted by Eq. (\ref{eq:44}). It can be observed that the change of reset time has a significant effect on the acquisition time when $U$ is small. The acquisition time with ${{T}_{a}}=30$ is larger than that with ${{T}_{a}}=10$ by $38\%$ at ${U}/{\kappa }=0.5$. The influence gradually weakens as FOU increases until ${U}/{\kappa }=2$ where the acquisition time with ${{T}_{a}}=30$ is only $2\%$ larger than that with ${{T}_{a}}=10$. This is because the number of resets decreases as $U$ increases, and the total reset time decreases. On the other hand, the ${T}_{U}$ is proportional to the square of $U$, which is a higher-order term relative to ${T}_{a}$. Moreover, the increase of the field detection probability ${P}_{V}$ means a larger single-scan acquisition probability, thus significantly reducing the acquisition time. Furthermore, the FOU was optimized with multi-scan as well in \citet{li2011analytical} and \citet{ma2021satellite}, obtaining the optimum FOU at ${U}/{\kappa }=1.3$, but the difference from ground-truth ${{U}_{opt}}$ is $1\%$ in the case of ${{T}_{a}}=30$ and ${{P}_{V}}=0.99$, while the difference reaches $38.3\%$ for ${{T}_{a}}=10$ and ${{P}_{V}}=0.9$. This is because that decreasing ${{T}_{a}}$ or ${{P}_{V}}$, which were not considered in \citet{li2011analytical} and \citet{ma2021satellite}, reduces ${{\hat{T}}_{a}}$ and further lower ${{U}_{opt}}$ according to Eq. (\ref{eq:43}). Additionally, the error of the fit ${{U}_{opt}}$ from Eq. (\ref{eq:44}) and the corresponding ${{T}_{M}}$ is within $0.2\%$ and ${{10}^{-4}}\%$, respectively, which demonstrates that our optimization conclusion of FOU is more accurate.

\begin{figure}[htbp!]
	\centering\includegraphics[width=0.5\columnwidth]{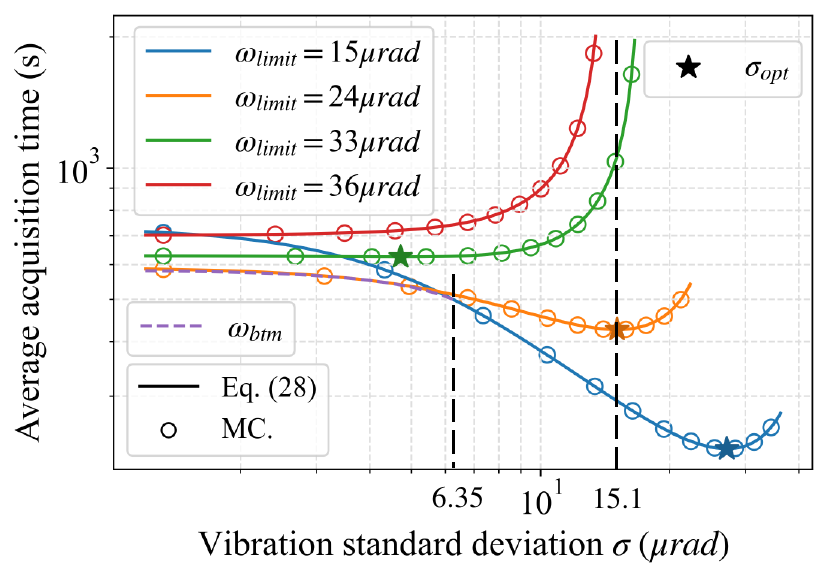}
	\caption{The variation of the multi-scan acquisition time with the platform vibration level under different beam divergence angles at turbulence level $Turb.3$.}
	\label{fig:7}
\end{figure}

Finally, Fig. \ref{fig:7} illustrates the variation of the multi-scan acquisition time with the platform vibration standard deviation under different beam divergence angles, where $d=80\mu rad$, $U=1.3mrad$, and turbulence level is $Turb.3$. When $\sigma <6.3\mu rad$, there is $B>B_{\sigma }^{\min }$ so that ${{\omega }_{btm}}$ exists, the corresponding acquisition time ${T}_{M}$ decreases with the increase of vibration level $\sigma $, which indicates that the vibration noise can be transformed into a favorable factor to improve the acquisition performance according to the optimization conclusion of divergence angle. For ${{\omega }_{limit}}=36\mu rad>{{B}^{{1}/{2}}}{{e}^{{-1}/{4}}}=34.3\mu rad$, the stronger the platform vibration, the longer the acquisition time. When $\sigma $ is close to $14.1\mu rad$, the coverage factor is approximately zero, resulting in the acquisition time approaching infinity. For ${{\omega }_{limit}}=\{15,24,33\}\mu rad$, the minimum ${T}_{M}$ exists at ${{\sigma }_{opt}}$, which increases with the decrease of ${{\omega }_{limit}}$. This shows that reducing ${{\omega }_{limit}}$ can improve the acquisition performance in an increasing range of platform vibration levels. When $\sigma \ge 6.3\mu rad$, there is $B\le B_{\sigma }^{\min }$ so that ${T}_{M}$ increases with ${{\omega }_{limit}}$ at the same vibration level $\sigma $. Moreover, for ${{\omega }_{limit}}=24\mu rad$, we obtain ${{\sigma }_{opt}}=15.1\mu rad$, where the corresponding acquisition time is increased by $134$ seconds and reduced by $640$ seconds compared with that for ${{\omega }_{limit}}=15\mu rad$ and ${{\omega }_{limit}}=33\mu rad$, respectively. However, reducing ${{\omega }_{limit}}$ means a larger resonator and a more complex system. Although both reduce ${{\omega }_{limit}}$ by $9\mu rad$, the cost from $24\mu rad$ to $15\mu rad$ is geometrically increased compared with that from $33\mu rad$ to $24\mu rad$. In other words, the improvement of the acquisition performance by reducing the beam divergence angle after ${{\omega }_{limit}}=24\mu rad$ is very limited at the same cost. This quantitatively demonstrates that designing ${{\omega }_{limit}}$ according to Eq. (\ref{eq:48}) achieves a good trade-off between the acquisition performance and complexity of the APT system.

\section{Conclusion}
In this paper, a multi-scan link acquisition time model based on the received average SNR is proposed for LEO-to-ground laser communication, where the parameters of beam divergence angle, spiral pitch, and FOU are optimized to obtain the minimum acquisition time. Such derivations had not been presented yet. Compared with the existing models that assumed a successful acquisition when the receiver is within the divergence angle of the Gaussian beam, the proposed model takes the received average SNR as the criterion, which is suitable for various photoelectric sensors. Specifically, we present the concept of "coverage factor", denoting the maximum ratio of acquisition angle to spiral pitch wherein the receiver meets SNR level for a certain transmitted power. Under the combined effects of the platform vibration with Rice distribution and the Gamma-Gamma turbulence channel, we derive the transmitted power as a function of beam divergence angle, spiral pitch, and coverage factor. For LEO-to-ground FSOC with a limited narrow window, the scanning needs to maintain the maximum transmitted power to achieve fast acquisition, thereby these parameters are decoupled at a fixed power. Subsequently, the probability distribution of coverage factor is derived based on the initial pointing error obeying Rayleigh distribution, which allows us to calculate the probability that the received SNR exceeds the threshold. Combined with FOU and receiver field angle, we obtain the single-scan acquisition probability, which is less than one, so that the multi-scan is adopted to ensure acquisition success. Considering the essential reset time between single-scan, we establish a novel multi-scan acquisition time model and present optimizations. The numerical results calculated by the proposed analytical expressions are consistent with the MC simulations. Due to the combination of turbulence, the proposed model is also applicable to the inter-satellite FSOC scenario. Moreover, the quantitative analysis of the influence of platform vibration indicates that the vibration noise will be transformed into a favorable factor to improve the acquisition performance in an increasing range of vibration levels as the decrease of the minimum divergence angle modulated by the laser. Furthermore, we present a theoretical method for designing the minimum divergence angle, which achieves a good trade-off between the link acquisition performance and complexity of the APT system. Overall, this work provides important theoretical support for the design of beaconless LEO-to-ground FSOC system.

\section*{Disclosures}
The authors declare no conflicts of interest.

\bibliographystyle{unsrtnat}
\bibliography{references}  

\begin{thebibliography}{31}
\providecommand{\natexlab}[1]{#1}
\providecommand{\url}[1]{\texttt{#1}}
\expandafter\ifx\csname urlstyle\endcsname\relax
  \providecommand{\doi}[1]{doi: #1}\else
  \providecommand{\doi}{doi: \begingroup \urlstyle{rm}\Url}\fi

\bibitem[Toyoshima et~al.(2007)Toyoshima, Leeb, Kunimori, and
  Takano]{toyoshima2007comparison}
Morio Toyoshima, Walter~R Leeb, Hiroo Kunimori, and Tadashi Takano.
\newblock Comparison of microwave and light wave communication systems in space
  applications.
\newblock \emph{Optical engineering}, 46\penalty0 (1):\penalty0 015003, 2007.

\bibitem[Toyoshima(2005)]{toyoshima2005trends}
Morio Toyoshima.
\newblock Trends in satellite communications and the role of optical free-space
  communications.
\newblock \emph{Journal of Optical Networking}, 4\penalty0 (6):\penalty0
  300--311, 2005.

\bibitem[Kim et~al.(2001)Kim, Riley, Wong, Mitchell, Brown, Hakakha, Adhikari,
  and Korevaar]{kim2001lessons}
Isaac~I Kim, Brian Riley, Nicholas~M Wong, Mary Mitchell, Wesley Brown, Harel
  Hakakha, Prasanna Adhikari, and Eric~J Korevaar.
\newblock Lessons learned for strv-2 satellite-to-ground lasercom experiment.
\newblock In \emph{Free-Space Laser Communication Technologies XIII}, volume
  4272, pages 1--15. SPIE, 2001.

\bibitem[Fields et~al.(2011)Fields, Kozlowski, Yura, Wong, Wicker, Lunde,
  Gregory, Wandernoth, and Heine]{fields20115}
Renny Fields, David Kozlowski, Harold Yura, Robert Wong, Josef Wicker, C~Lunde,
  Mark Gregory, B~Wandernoth, and F~Heine.
\newblock 5.625 gbps bidirectional laser communications measurements between
  the nfire satellite and an optical ground station.
\newblock In \emph{2011 International Conference on Space Optical Systems and
  Applications (ICSOS)}, pages 44--53. IEEE, 2011.

\bibitem[Gregory et~al.(2017)Gregory, Heine, K{\"a}mpfner, Meyer, Fields, and
  Lunde]{gregory2017tesat}
M~Gregory, F~Heine, H~K{\"a}mpfner, R~Meyer, R~Fields, and C~Lunde.
\newblock Tesat laser communication terminal performance results on 5.6 gbit
  coherent inter satellite and satellite to ground links.
\newblock In \emph{International Conference on Space Optics—ICSO 2010},
  volume 10565, pages 324--329. SPIE, 2017.

\bibitem[Young et~al.(1986)Young, Germann, and Nelson]{young1986pointing}
Philip~W Young, Lawrence~M Germann, and Roy Nelson.
\newblock Pointing, acquisition, and tracking subsystem for space-based laser
  communications.
\newblock In \emph{Optical technologies for communication satellite
  applications}, volume 616, pages 118--128. SPIE, 1986.

\bibitem[Picchi et~al.(1986)Picchi, Prati, and Santerini]{picchi1986algorithms}
G~Picchi, G~Prati, and D~Santerini.
\newblock Algorithms for spatial laser beacon acquisition.
\newblock \emph{IEEE transactions on aerospace and electronic systems},
  \penalty0 (2):\penalty0 106--114, 1986.

\bibitem[Yu et~al.(2017)Yu, Wu, Wang, Tan, and Ma]{yu2017theoretical}
Siyuan Yu, Feng Wu, Qiang Wang, Liying Tan, and Jing Ma.
\newblock Theoretical analysis and experimental study of constraint boundary
  conditions for acquiring the beacon in satellite--ground laser
  communications.
\newblock \emph{Optics Communications}, 402:\penalty0 585--592, 2017.

\bibitem[Hu et~al.(2022)Hu, Yu, Duan, Zhu, Cao, Zhou, Li, and Liu]{hu2022multi}
Siqi Hu, Hanghua Yu, Zheng Duan, Ye~Zhu, Caixia Cao, Miaomiao Zhou, Guotong Li,
  and Huijie Liu.
\newblock Multi-parameter influenced acquisition model with an in-orbit jitter
  for inter-satellite laser communication of the lces system.
\newblock \emph{Optics Express}, 30\penalty0 (19):\penalty0 34362--34377, 2022.

\bibitem[Ho(2007)]{ho2007pointing}
Tzung-Hsien Ho.
\newblock \emph{Pointing, acquisition, and tracking systems for free-space
  optical communication links}.
\newblock University of Maryland, College Park, 2007.

\bibitem[Hindman and Robertson(2004)]{hindman2004beaconless}
Charles Hindman and Lawrence Robertson.
\newblock Beaconless satellite laser acquisition-modeling and feasability.
\newblock In \emph{IEEE MILCOM 2004. Military Communications Conference,
  2004.}, volume~1, pages 41--47. IEEE, 2004.

\bibitem[Sterr et~al.(2011)Sterr, Gregory, and Heine]{sterr2011beaconless}
Uwe Sterr, Mark Gregory, and Frank Heine.
\newblock Beaconless acquisition for isl and sgl, summary of 3 years operation
  in space and on ground.
\newblock In \emph{2011 International Conference on Space Optical Systems and
  Applications (ICSOS)}, pages 38--43. IEEE, 2011.

\bibitem[Li et~al.(2011)Li, Yu, Ma, and Tan]{li2011analytical}
Xin Li, Siyuan Yu, Jing Ma, and Liying Tan.
\newblock Analytical expression and optimization of spatial acquisition for
  intersatellite optical communications.
\newblock \emph{Optics Express}, 19\penalty0 (3):\penalty0 2381--2390, 2011.

\bibitem[Friederichs et~al.(2016)Friederichs, Sterr, and
  Dallmann]{friederichs2016vibration}
Lothar Friederichs, Uwe Sterr, and Daniel Dallmann.
\newblock Vibration influence on hit probability during beaconless spatial
  acquisition.
\newblock \emph{Journal of Lightwave Technology}, 34\penalty0 (10):\penalty0
  2500--2509, 2016.

\bibitem[Ma et~al.(2021)Ma, Lu, Tan, Yu, Fu, and Li]{ma2021satellite}
Jing Ma, Gaoyuan Lu, Liying Tan, Siyuan Yu, Yulong Fu, and Fajun Li.
\newblock Satellite platform vibration influence on acquisition system for
  intersatellite optical communications.
\newblock \emph{Optics \& Laser Technology}, 138:\penalty0 106874, 2021.

\bibitem[Qiu et~al.(2021)Qiu, Lin, and Chen]{qiu2021active}
Zhaobing Qiu, Liyu Lin, and Liqiong Chen.
\newblock An active method to improve the measurement accuracy of four-quadrant
  detector.
\newblock \emph{Optics and Lasers in Engineering}, 146:\penalty0 106718, 2021.

\bibitem[Yang and Li(2022)]{yang2022iterative}
Sen Yang and Xiaofeng Li.
\newblock Iterative framework for a high accuracy aberration estimation with
  one-shot wavefront sensing.
\newblock \emph{Optics Express}, 30\penalty0 (21):\penalty0 37874--37887, 2022.

\bibitem[Li(2015)]{li2015limited}
Hanshan Li.
\newblock Limited magnitude calculation method and optics detection performance
  in a photoelectric tracking system.
\newblock \emph{Applied Optics}, 54\penalty0 (7):\penalty0 1612--1617, 2015.

\bibitem[Kaushal and Kaddoum(2016)]{kaushal2016optical}
Hemani Kaushal and Georges Kaddoum.
\newblock Optical communication in space: Challenges and mitigation techniques.
\newblock \emph{IEEE communications surveys \& tutorials}, 19\penalty0
  (1):\penalty0 57--96, 2016.

\bibitem[Hechenblaikner et~al.(2023)Hechenblaikner, Delchambre, and
  Ziegler]{hechenblaikner2023optical}
Gerald Hechenblaikner, Simon Delchambre, and Tobias Ziegler.
\newblock Optical link acquisition for the lisa mission with in-field pointing
  architecture.
\newblock \emph{Optics \& Laser Technology}, 161:\penalty0 109213, 2023.

\bibitem[Hemmati(2020)]{hemmati2020near}
Hamid Hemmati.
\newblock Near-earth laser communications.
\newblock In \emph{Near-Earth Laser Communications}, pages 1--40. CRC press,
  2020.

\bibitem[Gao et~al.(2023)Gao, Zou, Zhang, Wei, Kuang, and Zhu]{gao2023improved}
Geng Gao, Xiancai Zou, Shoujian Zhang, Hui Wei, Kaifa Kuang, and Kemin Zhu.
\newblock Improved real-time cycle-slip detection for low earth orbit
  satellites based on the dynamic force model.
\newblock \emph{Advances in Space Research}, 2023.

\bibitem[Steinhaus(1999)]{steinhaus1999mathematical}
Hugo Steinhaus.
\newblock \emph{Mathematical snapshots}.
\newblock Courier Corporation, 1999.

\bibitem[Riel et~al.(2020)Riel, Galffy, Janisch, Wertjanz, Sinn, Schwaer, and
  Schitter]{riel2020high}
Thomas Riel, Andras Galffy, Georg Janisch, Daniel Wertjanz, Andreas Sinn,
  Christian Schwaer, and Georg Schitter.
\newblock High performance motion control for optical satellite tracking
  systems.
\newblock \emph{Advances in Space Research}, 65\penalty0 (5):\penalty0
  1333--1343, 2020.

\bibitem[Andrews and Phillips(2005)]{andrews2005laser}
Larry~C Andrews and Ronald~L Phillips.
\newblock Laser beam propagation through random media.
\newblock \emph{Laser Beam Propagation Through Random Media: Second Edition},
  2005.

\bibitem[Jurado-Navas et~al.(2012)Jurado-Navas, Garrido-Balsells, Paris,
  Castillo-V{\'a}zquez, and Puerta-Notario]{jurado2012impact}
Antonio Jurado-Navas, Jos{\'e}~Mar{\'\i}a Garrido-Balsells, Jos{\'e}~Francisco
  Paris, Miguel Castillo-V{\'a}zquez, and Antonio Puerta-Notario.
\newblock Impact of pointing errors on the performance of generalized
  atmospheric optical channels.
\newblock \emph{Optics Express}, 20\penalty0 (11):\penalty0 12550--12562, 2012.

\bibitem[Toyoshima et~al.(2002)Toyoshima, Jono, Nakagawa, and
  Yamamoto]{toyoshima2002optimum}
Morio Toyoshima, Takashi Jono, Keizo Nakagawa, and Akio Yamamoto.
\newblock Optimum divergence angle of a gaussian beam wave in the presence of
  random jitter in free-space laser communication systems.
\newblock \emph{JOSA A}, 19\penalty0 (3):\penalty0 567--571, 2002.

\bibitem[Rice(1948)]{rice1948statistical}
Stephen~O Rice.
\newblock Statistical properties of a sine wave plus random noise.
\newblock \emph{The Bell System Technical Journal}, 27\penalty0 (1):\penalty0
  109--157, 1948.

\bibitem[Al-Habash et~al.(2001)Al-Habash, Andrews, and
  Phillips]{al2001mathematical}
Ammar Al-Habash, Larry~C Andrews, and Ronald~L Phillips.
\newblock Mathematical model for the irradiance probability density function of
  a laser beam propagating through turbulent media.
\newblock \emph{Optical engineering}, 40\penalty0 (8):\penalty0 1554--1562,
  2001.

\bibitem[Wang and Cheng(2010)]{wang2010moment}
Ning Wang and Julian Cheng.
\newblock Moment-based estimation for the shape parameters of the gamma-gamma
  atmospheric turbulence model.
\newblock \emph{Optics express}, 18\penalty0 (12):\penalty0 12824--12831, 2010.

\bibitem[Proke{\v{s}}(2009)]{prokevs2009modeling}
Ale{\v{s}} Proke{\v{s}}.
\newblock Modeling of atmospheric turbulence effect on terrestrial fso link.
\newblock \emph{Radioengineering}, 18\penalty0 (1):\penalty0 42--47, 2009.

\end{thebibliography}






\end{document}